\documentclass[a4paper,12pt]{article}

\usepackage{amssymb,amsmath}
\usepackage[pdftex]{graphicx}
\usepackage[utf8]{inputenc}
\title{Fault Tolerance of Small-World Regular and Stochastic Interconnection Networks}
\author{A.~Demichev$^{1}$, V.Ilyin$^{1,2}$, A. Kryukov$^{1}$ and S.Polyakov$^1$\\
{\small\textit{$^1$Skobeltsyn Institute of Nuclear Physics, Lomonosov Moscow State University}}\\ 
{\small\textit{$^2$National Research Centre "Kurchatov Institute"}}}
\date{}

\begin{document}

\maketitle
\begin{abstract}
Resilience of the most important properties of stochastic and regular (deterministic) small-world interconnection networks is studied. It is shown that in the broad range of values of the fraction of faulty nodes the networks under consideration possess high fault tolerance, the deterministic networks being slightly better than the stochastic ones. 

\end{abstract}

\section{Introduction\label{Vved}}

One of the most important components of any supercomputer is the interconnection network, which primarily defines the ability to increase the number of computing nodes necessary for achieving the desirable performance. According to very modest estimations (for example, see \cite{INW}), interconnection networks of Exaflop supercomputers must have more than 100 000 endpoints/CPUs (each most likely with many cores). Thus, one of the key problems, which should be solved on a way to next-generation supercomputers, is the development of an interconnection network architecture with good scalability and the ability to provide communication between a huge number of computing nodes.

Among direct networks, in which every node serves both as a processing unit and as a router, the networks with lattice structure and $D$-dimensional torus topology are widely used. However, in the case of the large number of nodes typical for next-generation supercomputers, the $D$-dimensional lattice toruses have essential shortcomings. In particular, lattices of low dimensionality have very large average path length between nodes while lattices of high dimensionality are inconvenient to embed into real physical space and therefore for implementation because of the large length of communication links. For this reason it seems attractive to use the interconnection  lattice networks with additional links (shortcuts) between the nodes. In the case of an appropriate construction such shortcuts provide small average distance $d$ between nodes and the small diameter ${\mathcal D}$ of the networks (small-world networks \cite {WS}). Such networks can be constructed  by means of both deterministic and stochastic algorithms (see, for example, the paper \cite {DIKP} and refs therein). For many real networks the stochasticity is an inherent property (e.g., this is true for the Web and the Internet). However, in the case of interconnection networks the design/construction process is under full control of a developer. Therefore, the stochasticity is not inherent for this process. Thus an important question is the following: is there such a deterministic algorithm for lattice modification turning it into a small-world network that the cost-quality ratio is better than in the case of stochastic algorithms. In the work \cite {DIKP} a comparison of a number of deterministic and stochastic algorithms was carried out from the point of view of the cost-quality ratio optimization, the ``cost'' being total shortcuts length per node (the unit wiring cost) while ``quality'' being the global or navigation (see \cite{K2000}) average path length between nodes. It has been shown that there exist deterministic networks having the cost-quality ratio equal or better than that for stochastic networks.

However, for  very-large-scale interconnection networks, which is expected to be typical for Exaflop supercomputers, faults of a part of nodes are almost inevitable because of physical failures or simply of an overloads. Therefore the fault tolerance of such networks is an important issue. The failures of a part of nodes  results in some stochasticity even for regular interconnection networks constructed by deterministic algorithms. Therefore the node failures may affect the regular networks stronger than the stochastic ones.

This work aims at the comparison of fault tolerance ability of deterministic (regular) and stochastic (irregular) networks. To achieve this we consider networks on the basis of two-dimensional ($D=2$) regular lattice with shortcuts added by means of a stochastic or deterministic algorithms. Only topological characteristics such as average navigation distance, forwarding index as well as possible cascading failures are considered. Dynamical models of message transfers in the interconnection networks, in particular on the basis of the theory of queuing systems, will be studied in the subsequent works.

\section{Algorithms for Shortcuts Constructing and Message Routing\label{VSDS}}

We will use the following version of the stochastic algorithms (comparison of various algorithms see in \cite{DIKP}):
primary object is the two-dimensional lattice with $N=L\times L$ nodes and topology of the two-dimensional torus; one goes sequentially through all the network nodes and attach the first end of a shortcut to each node $(i,j),\ 1\leq i,j\leq L$; the second shortcut end (that is the node $(k,l)$ to which the end is attached) must not coincide with the neighbors of the first end in the sense of the underlying lattice and must not result in the duplication of an existent shortcut; in all other aspects it is chosen randomly with a probability $P(r)\sim r^{-\alpha}$ which is the power-like function of the lattice distance $r=r_{(ij),(kl)}$ between the nodes $(i,j)$ and $(k,l)$ ($r$ denotes the lattice distance between arbitrary couple of nodes). The dependence of the probability on the distance between the nodes reflects correlation between topological and spatial network properties \cite{Bar}.

Among variety of deterministic networks with small average distance between nodes, the Interlaced Bypass Torus Networks (iBT networks) \cite{ZPD} possess the most preferable general characteristics. Therefore we will compare the fault tolerance ability of this deterministic network with the stochastic ones. The detailed description and the analysis of these networks can be found in \cite{ZD}.

As a matter of rule, for routing messages in large networks adaptive algorithms (see, for example, \cite {DT}) are used. However, efficiency of a specific adaptive algorithm significantly depends on type and architecture of a network for which it is used.  Inasmuch as in this work we compare essentially different interconnection networks (stochastic and deterministic), we use the most general adaptive algorithm of routing based on the principles of local navigation \cite{K2000}. In the case of local navigation a node ``knows'' only geographical locations of all other nodes (in other words, their locations in the lattice) as well as its nearest neighbors with account of the shortcuts. Using only such information (the information about all the shortcuts in the network is not used) it is necessary to deliver the message to the target node along possibly shortest path. The simplest solution to this problem is to use the greedy algorithm: the current node sends the message to that of its neighbors which geographically (in the sense of coordinates on the underlying lattice) is closest to the destination node. 

However, some networks possess rather large average path length in the case of local navigation based on the elementary greedy algorithm. To improve the situation, one may use the modified algorithm of local navigation, namely the algorithm of two-level local navigation \cite{DIKP}. In this modification, not only the nearest neighbors of a given node are taken into account, but also neighbors of neighbors: a message from a given node is transferred to its neighbor which, in turn, has a neighbor closest to the destination node in the sense of the underlying lattice. The two-level greedy algorithm requires a little bit more calculations at each node but still remains local and well scalable. The average path length in case of the two-level local navigation is denoted by $\ell^{(2)}$. Notice that the global average path length $d$ can be considered as a limiting case of navigation length with an infinite depth of search.

For a regular network, if a routing bound to the specific regular structure is used, in case of nodes failures (and hence destruction of this regular structure) path length between nodes would change stronger, than in the case of the greedy algorithms which are universal ones for any kind of networks. In other words, the greedy algorithms might be not optimal for a specific  regular network, but the steadiest one with respect to failure of a part of network nodes.

Certainly, in the case of possibility of failure of part of nodes the greedy algorithm requires more precise definition when the node to which the message should be transfered proves to be faulty. In particular, it is forbidden to return to a node from which the message came at the previous step. The message is considered lost, if: there is no place to go to; or the length of the path covered exceeds $2{\mathcal D}^{(2)}$, where ${\mathcal D}^{(2)}$ is the navigation diameter, i.e. the maximum path length in a network for the two-level local navigation.

\section{Behaviour of the Networks under Node Failures\label{USVSU}}

The fault tolerance comparison was accomplished for best samples of the interconnection networks of each type having best values of average navigation path length $\ell^{(2)}$ and forwarding index (maximum number of paths passing through any node in the network; see, e.g., \cite{Xu}) \textit {without} the failed nodes. Nodes which are considered to be failed were selected randomly, several times for each copy of a network (10 for the provided plots) and for each value $b$ of number of faulty nodes. The results were averaged over the selections.

\begin{figure}
\begin{center}
\includegraphics[scale=.6]{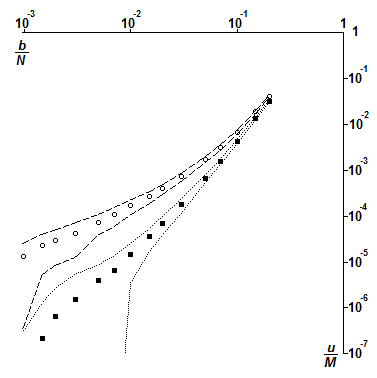}
\end{center}
\caption{Dependence of a fraction $u/M$ of the undelivered messages on a fraction of the failed nodes in the stochastic networks (circles, $\circ$) and in the iBT-networks (filled squares, $\scriptstyle\blacksquare$); the dashed and the dotted curves show the ranges of the root-mean-square deviations
\label{fig:u_of_b_L128_S1m_alpha_1_and_iBT}}
\end{figure}

\begin{figure}
\begin{center}
\includegraphics[scale=.6]{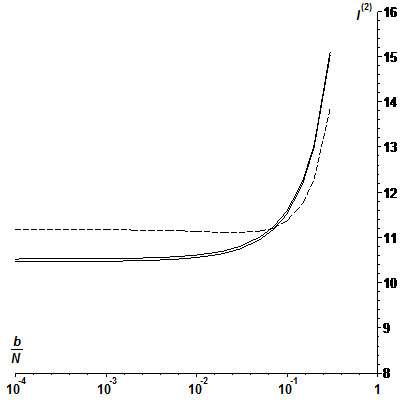}
\end{center}
\caption{Dependence of the average navigation distance $\ell^{(2)}$ on a fraction $b/N$ of failed nodes in the stochastic networks (solid lines) and in the iBT-networks in the case of shortcuts lengths $\{8,\ 32\}$ (broken line); the linear size of the networks is $L=128$
\label{fig:l2_of_b_S1m_iBT}}
\end{figure}

\begin{figure}
\begin{center}
\includegraphics[scale=.7]{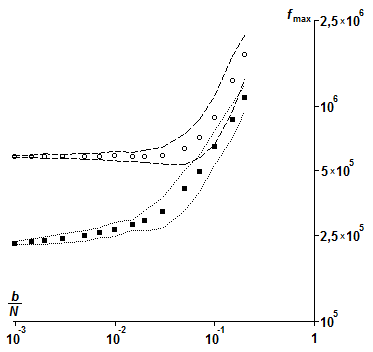}
\end{center}
\caption{Dependence of the forwarding index $f_{max}$ on a fraction $b/N$ of failed nodes in the stochastic networks (circles, $\circ$) and in the iBT-networks (filled squares, $\scriptstyle\blacksquare$); the dashed and the dotted curves show the ranges of the root-mean-square deviations
\label{fig:f_max_of_b_plain_lattice_S1m_iBT}}
\end{figure}

\begin{figure}[t]
\begin{center}
\includegraphics[scale=.55]{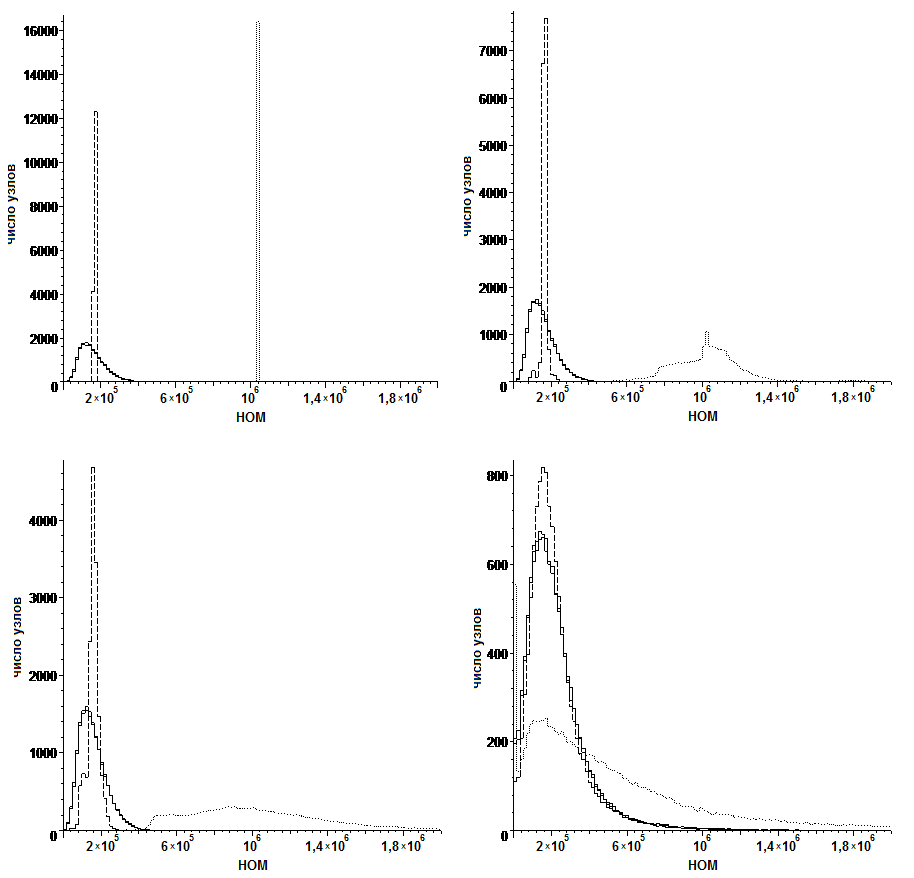}
\end{center}
\caption{Distribution of loadings with respect to the routing algorithm (two-level greedy algorithm for the stochastic networks (solid lines), iBT  (broken line) and the torus without shortcuts (dotted line) in the cases of fractions of failed nodes (from left to right and from top to down): $b=0,\ 0,01N,\ 0,05N,\ 0,3N$; $L=128$
\label{fig:f_distribution_with_torus}}
\end{figure}

\begin{figure}
\begin{center}
\includegraphics[scale=.7]{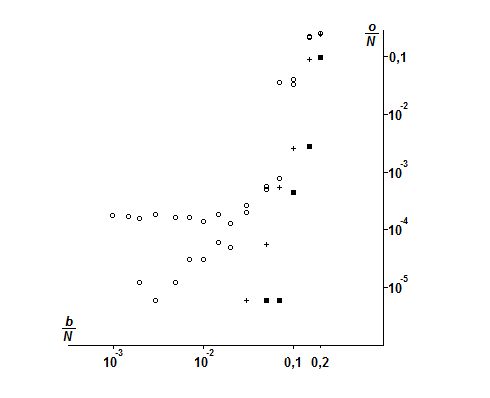}
\end{center}
\caption{Dependence of a fraction of overloaded nodes on a fraction of failed nodes in the stochastic networks (circles, $\circ$), in stochastic networks with degree of all the nodes equal to six (plus signs, +) and in the iBT-networks (filled squares, $\scriptstyle\blacksquare$); the maximum permissible load is equal $3f_{max}^{(iBT)}$
\label{fig:o_of_b_L_128_S1m_alpha_1_SR1_alpha_1_and_iBT_max_f_512907_log_scale}}
\end{figure}

The simulation results for resilience of a number of features for stochastic and deterministic networks are presented in fig.~\ref{fig:u_of_b_L128_S1m_alpha_1_and_iBT} -- \ref{fig:o_of_b_L_128_S1m_alpha_1_SR1_alpha_1_and_iBT_max_f_512907_log_scale}:
\begin {itemize}
 \item dependence of a fraction $u/M$ of the undelivered messages ($M$ is the total number of messages) on the fraction $b/N$ of the failed nodes (fig.~\ref{fig:u_of_b_L128_S1m_alpha_1_and_iBT});
 \item dependence of the average navigation distance $\ell^{(2)}$ on $b/N$ (fig.~\ref {fig:l2_of_b_S1m_iBT});
 \item dependence of the forwarding index $f_{max}$ on $b/N$ (fig.~\ref {fig:f_max_of_b_plain_lattice_S1m_iBT});
 \item distribution of loadings with respect to the routing algorithm (two-level greedy algorithm; fig.~\ref {fig:f_distribution_with_torus});
 \item dependence of the fraction of the overloaded nodes on $b/N$ (fig.~\ref{fig:o_of_b_L_128_S1m_alpha_1_SR1_alpha_1_and_iBT_max_f_512907_log_scale}).
\end{itemize}

The results are provided for networks with the best values of $\ell^{(2)}$: in the stochastic case the networks are selected out of 100 samples with parameter $\alpha=1$; in the case of iBT-networks the best variants with two lengths of shortcuts (namely, $s_1=8$, $s_2=32$) were selected. As concerns the chosen value of the parameter $\alpha$, it is worth making the following remark. In the well-known work \cite{K2000} it is shown that in case of two-dimensional networks average navigation length most slowly (logarithmically) grows with increasing in the sizes of a network in case of $\alpha=2$. However, this result is inapplicable directly to the case under consideration because:
\begin {itemize}
 \item in the paper \cite{K2000} the simple greedy algorithmis is considered while in our work its two-level generalization is used; 
 \item the results of the paper \cite {K2000} provide the asymptotical behavior of the average navigation length for growing network size while for specific finite value of network size the optimal value of the parameter $\alpha$ may significantly differ from two. 
\end {itemize}
The simulations show that for the chosen network size ($L=128$; $N=L^2=16384\approx 1.6\times 10^4$) the networks with $\alpha\approx1$ have the smallest average navigation lengths. 

When evaluating the average navigation distance $\ell^{(2)}$, the undelivered messages were not taken into account. Obviously, this results in a lower bound on the value of $\ell^{(2)}$ shown in fig.~\ref {fig:l2_of_b_S1m_iBT}. To compare different networks, some value should be assigned to the paths covered by undelivered messages. However, the simulations show that the qualitative behavior of the results does not depend on the specific value of the path length assigned to the undelivered messages if this path length $\gtrsim 2{\mathcal D}^{(2)}$ and practically matches the lower estimation presented in fig.~\ref {fig:l2_of_b_S1m_iBT}. The forwarding index produced by undelivered messages is calculated in the same way as that produced by successfully delivered messages. 

When a part of nodes fails, the loading of the remaining nodes increases. If for some node this increased loading reaches the threshold, namely the maximum permissible load for this node, this node actually too ceases to be operable. Thus the cascade process of switching-off of the nodes can progress (see, for example, \cite {ML}, \cite {DCN}). To model such a situation, for all the nodes of the networks a limiting load (threshold) $f_ {th}$ is defined above which a node ceases to transfer messages, actually it fails. The cascading failure in this model looks as follows. Some initial fraction of nodes chosen randomly is removed from the network (it is supposed that these nodes failed owing to the hardware failures or an accidental overload not caused by reallocation of load). This leads to reallocation of load to other nodes. All the overloaded nodes (i.e., with the load higher than $f_ {th}$)  are determined and  removed from the network. The latter again leads to reallocation of load and the procedure is repeated while there exist overloaded nodes.

The dependence of the fraction of the overloaded nodes on the fraction of the nodes made inoperative originally in stochastic and iBT-networks is shown in fig.~\ref {fig:o_of_b_L_128_S1m_alpha_1_SR1_alpha_1_and_iBT_max_f_512907_log_scale}. The maximum permissible load $f_{th}$ is defined on the basis of the load distribution in nonfaulty small-world networks (the upper left plot in the fig.~\ref{fig:f_distribution_with_torus}). Namely, $f_ {th}$ is chosen to be a multiple of the maximum value of loading for iBT-networks: $f_{th}= k f_{max}^{(iBT)}$ (i.e., under designing the interconnection networks, the routers are selected with assurance factor $k$ relative to the iBT-networks). Simulations shows that if to choose $k=2$, the part of nodes in the stochastic networks proves to be immediately overloaded and this overload causes the cascading failure for all the network.  The results for $k=3$ are presented in fig.~\ref{fig:o_of_b_L_128_S1m_alpha_1_SR1_alpha_1_and_iBT_max_f_512907_log_scale}.

\section{Results and Conclusion\label{Zak}}

In the case of rather small fraction of faulty nodes the amount of lost messages proves to be smaller in the deterministic iBT-networks, while when the fraction of nodes approaches $\sim 10\%$, the ratio of lost messages is practically the same for the both types of the networks. From the point of view of resilience of average navigation distance $\ell^{(2)}$ between nodes, the stochastic and iBT-networks behave approximately equally: till the fraction of faulty nodes $\lesssim 10\%$, the average navigation path length practically does not change and when the fraction $\gtrsim 10\%$ the average navigation path length sharply increases. Similarly to the average navigation distance, the forwarding index almost does not change till the fraction of faulty nodes does not reach  $\approx 10\%$ and both types of the networks (stochastic and deterministic) behave approximately equally (though loading for iBT-networks is slightly lower). After the fraction reaches $\gtrsim 10\%$, the maximum load $f_{max}$ sharply increases.

Load distribution in the stochastic networks is predictably significantly wider, than in the iBT-networks, the forwarding index in the former being approximately twice higher, than in the latter. It means that the throughput of routers in the stochastic networks must be essentially higher than in the iBT-networks for equal sizes of the networks. The load distribution in the iBT-networks remains narrower up to the value of the fraction of faulty nodes  $\sim 20\%~\div~30\%$. Fixing the node degree (i.e., the number of links entering a node) in the stochastic networks slightly reduces the width of the distribution of loadings, however the stochasticity in locations of the shortcuts still causes significantly different loading of the network nodes.

Comparing of the load distribution for the small-world networks with that in case of an ordinary lattice with torus topology is quite indicative: for completely  nonfaulty lattice without shortcuts the load on each node is identical but significantly higher, than for the small-world networks (as a result of significantly higher value of average distance between nodes for a usual lattice). In the case of failure of even a small fraction of the nodes ($\sim 1\%\ \div\ 5\%$) the load distribution width sharply increases and so a considerable fraction of the nodes may experience essential overloads. Thus the small-world networks not only provide a smaller time delay for message transfers because of small average distance between nodes, but also have significantly higher fault tolerance ability.

Our study of cascading failures shows that the both types of networks behave approximately equally: up to the values of a fraction of failed nodes $\sim 10\%\ \div\ 15\%$ the cascading failures do not occur, while above these values there is sharp growth of overloaded, and as a result, the failed nodes. With that the iBT-networks are a little bit steadier against the cascading failures than the stochastic networks. Fixing the number of links entering into a node in the stochastic networks has no essential impact on the cascade processes.

As a whole the results of the present work show that unlike the usual lattices with torus topology, the studied small-world networks in the broad range of values of the fraction of failed nodes have rather high fault tolerance ability. The iBT-networks constructed on the basis of the deterministic algorithm behave slightly better than the stochastic networks. 

\section*{Acknowledgments}

This work is partially supported by RFBR grant No. 12-07-00408-a.

\end{document}